\documentstyle{mn}
\input psfig

\newcount\eqnno \eqnno=0 
\def\numeqn{\global\advance\eqnno by 1 \eqno(\the\eqnno)}

\title[The Generalized Poisson distribution]
{The Generalized Poisson distribution and a model of clustering 
from Poisson initial conditions}

\author[R. K. Sheth]
{Ravi K. Sheth\\
Max-Planck Institut f\"ur Astrophysik, Karl-Schwarzschild-Str. 1,
85740 Garching, Germany\\
\smallskip
Email: sheth@mpa-garching.mpg.de
}

\date{Accepted 1998 March.  
Received 1997 November; in original form 1997 August}

\begin{document}

\maketitle

\begin{abstract}
This paper presents a new derivation of the Generalized Poisson 
distribution.  The derivation is based on the barrier crossing 
statistics of random walks associated with the Poisson 
distribution.  A simple interpretation of this model in terms 
of a single server queue is also included.  

In the astrophysical context, the Generalized Poisson 
distribution is interesting because it provides a good fit to 
the evolved, Eulerian counts-in-cells distribution measured in 
numerical simulations of hierarchical clustering from Poisson 
initial conditions.  The new derivation presented here can be 
used to construct a useful analytic model of the evolution of 
clustering measured in these simulations.  The model is consistent 
with the assumption that, as the universe expands and the comoving 
sizes of regions change as a result of gravitational instability, 
the number of such expanding and contracting regions is conserved.  
The model neglects the influence of external tides on the evolution 
of such regions.  Indeed, in the context of this model, 
the Generalized Poisson distribution can be thought of as arising 
from a simple variant of the well-studied spherical collapse model, 
in which tidal effects are also neglected.  This has the following 
implication:  Insofar as the Generalized Poisson distribution derived 
from this model is a reasonable fit to the numerical simulation 
results, the counts-in-cells statistic must be relatively insensitive 
to such effects.  This may be a consequence of the Poisson initial 
condition.  

The model can be understood as a simple generalization of the 
excursion set model which has recently been used to estimate 
the number density of collapsed, virialized halos.  The 
generalization developed here allows one to estimate the evolution 
of the spatial distribution of these halos, as well as their number 
density.  For example, it provides a framework within which the 
halo--halo correlation functions, at any epoch, can be computed 
analytically.  In the model, when halos first virialize, they are 
uncorrelated with each other.  This is in good agreement with the 
simulations.  Since it allows one to describe the spatial 
distribution of the halos and the mass simultaneously, the model 
allows one to estimate the extent to which these halos are biased 
tracers of the underlying matter distribution.  

\end{abstract} 
\maketitle

\section{Introduction}
Consider an initially Poisson distribution of particles that 
clusters gravitationally as the universe expands.  In this paper, 
the initial Poisson distribution will also be called the initial 
Lagrangian distribution.  As time passes, the particle distribution 
evolves, as, for example, tightly bound virialized clusters 
(called halos, or dark matter halos, in this paper) form.  
Thus, the evolved distribution is different from the initial 
Lagrangian distribution.  In what follows, the evolved distribution 
will be called the Eulerian distribution.  The goal of this 
paper is to use the properties of the initial Lagrangian distribution 
to derive a reasonable approximation to the form of the evolved 
Eulerian distribution.  
In the absence of a model relating the two distributions, the only 
constraint is that required by mass conservation:  the number of 
particles in the initial and evolved distributions is the same, 
so the average density, $\bar n$, in the two distributions must be 
the same.  In what follows, quantities measured in the Lagrangian 
space will have a subscript `0', while those in Eulerian space will 
not.  In this notation, mass conservation implies that 
$\bar n_0=\bar n$.  

Studies of clustering from Poisson initial conditions 
(Itoh, Inagaki \& Saslaw 1993 and references therein) show that 
when the initial, Lagrangian distribution is Poisson, then the 
evolved Eulerian distribution is Generalized Poisson.  This paper 
presents a model in which this is so.  The model is consistent 
with three general hypotheses about the evolution of clustering.  
The first is the hypothesis that, in comoving coordinates, 
initially denser regions contract more rapidly than less dense 
regions, and that sufficiently underdense regions expand.  
The second assumption is that, as the universe evolves, the number 
of such expanding and contracting regions is conserved---only 
their comoving size changes.  The third is that the influence of 
external tides on the evolution of such comoving regions can be 
neglected, if one is only interested in computing statistics such 
as the mass function of collapsed halos, or the distribution 
of counts in Eulerian cells.  
There are no compelling physical arguments for any of these 
assumptions, and initial particle configurations which violate 
some or all of these assumptions are relatively easy to construct.  
That the model predicts a counts-in-cells distribution which is a 
reasonable approximation to that measured in the numerical 
simulations suggests that, at least for clustering from Poisson 
initial conditions, these simple assumptions may also be 
reasonably accurate.  

\subsection{The Generalized Poisson distribution}\label{gpdsec}
Since it plays a central role in this paper, various 
known properties of the Generalized Poisson distribution are 
summarized below.  

The Generalized Poisson distribution (Consul 1989) has the form 
\begin{equation}
p(N|V,b) = {\bar N(1-b)\over N!}\,
\Bigl[\bar N(1-b) + Nb\Bigr]^{N-1} {\rm e}^{-\bar N(1-b) - Nb}\!.
\label{gpdf}
\end{equation} 
Here $p(N|V,b)$ is the probability that 
a cell of size $V$ placed randomly within a particle distribution 
contains exactly $N$ particles.  If $\bar n$ denotes the average 
density, then $\bar N\equiv \bar nV$.  In this paper $0\le b<1$, 
and, for reasons discussed below, it will be supposed that $b$ 
is not a function of $V$.  The case $b=0$ is the Poisson 
distribution.  

Equation~(\ref{gpdf}) is a Compound Poisson distribution 
(e.g. Saslaw 1989); it arises if point sized clusters, 
called halos in the following, have a Poisson spatial distribution, 
and the probability a randomly chosen halo contains exactly $n$ 
particles is 
\begin{equation}
\eta(n,b) = {(nb)^{n-1}\,{\rm e}^{-nb}\over n!}.  
\label{borel}
\end{equation}
This is the Borel$(b)$ distribution (Borel 1942).  In this 
paper, equation~(\ref{borel}) will be called the halo mass function.  

The Generalized Poisson distribution was first discovered in the 
astrophysical context by Saslaw \& Hamilton (1984) (also see 
Sheth 1995a).  It provides a good fit to the distribution of 
particle counts in randomly placed cells, provided the particle 
distributions have evolved, as a result of gravitational clustering, 
from an initially Poisson distribution (Itoh, Inagaki \& Saslaw 1993 
and references therein).  In fact, the fits are significantly 
improved if $b$ is allowed to increase to an asymptotic value as 
$V$ increases.  This scale dependence is simply a consequence of 
relaxing the assumption that Borel clusters are point sized, but 
still requiring that they have some finite size.  The asymptotic 
value of $b$ is that which would have characterized the distribution, 
had the clusters been point sized (Sheth \& Saslaw 1994).  
For this reason, the asymptotic value of $b$ is fundamental, and 
the point sized idealization useful.  This paper is mainly concerned 
with the point sized idealization, so that, in what follows, $b$ 
is independent of $V$.  

The point sized idealization is also motivated by the following 
observation.  To a good approximation, the distribution of bound 
virialized halos in the numerical simulations is Borel$(b)$.  
Thus, to a good approximation, clustering from Poisson initial 
conditions evolves in such a way that, at all times, particles 
are bound up in Borel$(b)$ halos, and, at the time when they first 
virialize, these halos have a Poisson distribution.  The evolution 
of clustering is parameterized by the time dependence of $b$; it is 
zero initially, and it increases as the universe expands (Zhan 1989; 
Sheth 1995b and references therein).  Therefore, in the remainder 
of this paper, $b$ will be treated as a pseudo-time variable, and 
the Borel$(b)$ distribution will often be called the halo mass 
function at the epoch $b$.  

As $V\to 0$, most cells in the Lagrangian and Eulerian distributions 
will be empty.  Equation~(\ref{gpdf}) shows 
that, in this limit, the probability that a cell is not empty is 
$\bar N(1-b)$, and the probability that a non-empty cell contains 
exactly $N$ particles is given by the Borel$(b)$ distribution.  
In other words, at the epoch $b$, the halo mass function is the 
same as the vanishing-cell-size limit of the Eulerian counts in 
cells distribution (Sheth 1996a).  This fact will be useful later.  

The Borel$(b)$ distribution can be derived from a number of 
different constructions, all of which are related to the Poisson 
distribution (Epstein 1983; Sheth 1995b; Sheth 1996b; Sheth \& 
Pitman 1997).  In the context of this paper, all these constructions 
can be thought of as providing models that allow one to compute the 
Eulerian space distribution, in the limit of vanishing cell size, 
given that the Lagrangian space distribution is Poisson.  
One of these constructions, based on the statistics of random 
walk  barrier crossings associated with the Poisson distribution, 
is the excursion set model (Epstein 1983; Sheth 1995b).  
This paper shows how to derive the Generalized Poisson distribution 
from a simple generalization of this excursion set model.  
The generalization shows how to derive the Eulerian space 
Generalized Poisson distribution from the Lagrangian space 
Poisson distribution, for all cell sizes, and all times.  

\subsection{Outline of this paper}
Section~\ref{const} summarizes the random walk, excursion set 
model which leads to the Borel$(b)$ distribution.  
Sections~\ref{shift} and~\ref{expdf} describe a generalization of 
this model which leads to a new derivation of the Generalized 
Poisson distribution.  
Section~\ref{halos} shows how to describe the spatial distribution 
of virialized halos within the context of this model.  It shows that 
the model is consistent with the Compound Poisson interpretation 
of the Generalized Poisson distribution -- in the model, Borel$(b)$ 
halos have a Poisson distribution at the time when they first 
virialize.  Moreover, in the model, the $V\to 0$ limit of the 
counts-in-cells distribution is, indeed, the halo mass function.  
This shows explicitly that the excursion set approach developed here 
is able to reproduce the known properties of the Generalized Poisson 
distribution.  The relation between this model and the well-studied 
spherical collapse model (outlined in Appendix~\ref{scoll}) is 
discussed in Section~\ref{scm}.  

Section~\ref{queue} contains a brief digression which relates 
the excursion set model of the previous section to a simple 
single server queue system.  Section~\ref{scale} discusses 
a scaling solution associated with the model that is analogous 
to the scaling solution found in Section~3.2 of Sheth (1995b).  
Section~\ref{twobar} discusses the associated two barrier 
problem.  The solution of this problem may provide 
useful diagnostics in assessing the rate of evolution of the 
Eulerian statistics computed earlier in the paper.  

Clustering from more general initial conditions, using the 
techniques developed here, is treated in a forthcoming paper.  

\section{Poisson initial conditions and the Generalized Poisson 
distribution}\label{picgpd}
This section presents a new derivation of the Generalized Poisson 
distribution.  The derivation uses a simple generalization of the 
excursion set model studied by Epstein (1983) and Sheth (1995b).  

\subsection{The excursion set with constant barrier}\label{const}
Suppose that the initial Lagrangian distribution is Poisson, 
with mean density $\bar n$.  This means that a volume of size 
$V_0$ placed at a random position in this distribution will 
contain exactly $N$ particles with probability 
\begin{equation}
p(N|V_0) = {\bar N_0^N\,{\rm e}^{-\bar N_0}\over N!} ,\qquad
{\rm where}\ \bar N_0=\bar nV_0.
\label{pois}
\end{equation}
Now choose a random particle of this distribution, and compute 
the density within concentric spheres centred on this position.  
Call the curve $N(V_0)$ traced out by the number of particles 
contained within a sphere $V_0$ centred on this particle, as a 
function of the sphere size $V_0$, a trajectory.  Then each 
particle in the Poisson distribution has its associated Lagrangian 
trajectory.  Given $\delta_{\rm c0}$, Epstein (1983) derived 
an expression for the fraction of Lagrangian trajectories for which 
$N(V_0) = \bar nV_0(1+\delta_{\rm c0})$, and for which 
$N(V_0') < \bar nV_0'(1+\delta_{\rm c0})$ for all $V_0'>V_0$ 
(also see Sheth 1995b; Sheth \& Lemson 1998).  

Epstein argued that a given value of $\delta_{\rm c0}\ge 0$ 
defines a series of volumes $V_0(1)< V_0(2)< \ldots$ for which  
\begin{equation}
j/V_0(j) = \bar n(1+\delta_{\rm c0}) \equiv \bar n/b,  
\qquad{\rm where}\ 0\le b<1.
\label{vj}
\end{equation}
The final equality defines $b = 1/(1+\delta_{\rm c0})$.  
The evolution of clustering enters through the time 
dependence of $\delta_{\rm c0}$, which decreases as 
time increases.  It is in this sense that $b$ is a 
pseudo-time variable; it is 0 initially, and increases as the 
universe expands (Sheth 1995b).  

Epstein showed that the fraction of trajectories 
$f_{\rm c}(j,b_1)$ for which $V_0(j)$ is the largest value of 
$V_0$ at which $N(V_0)=\bar N_0/b_1$ is 
\begin{equation}
f_{\rm c}(j,b_1) = 
(1-b_1)\,{(jb_1)^{j-1}\,{\rm e}^{-jb_1}\over (j-1)!}
= j\,(1-b_1)\,\eta(j,b_1),
\label{fjb}
\end{equation}
where $\eta(j,b)$ is the Borel$(b)$ distribution defined earlier.  
The mean of the Borel$(b)$ distribution is  
$\sum j\,\eta(j,b)=(1-b)^{-1}$, a fact which will be useful later.  

\begin{figure}
\centering
\mbox{\psfig{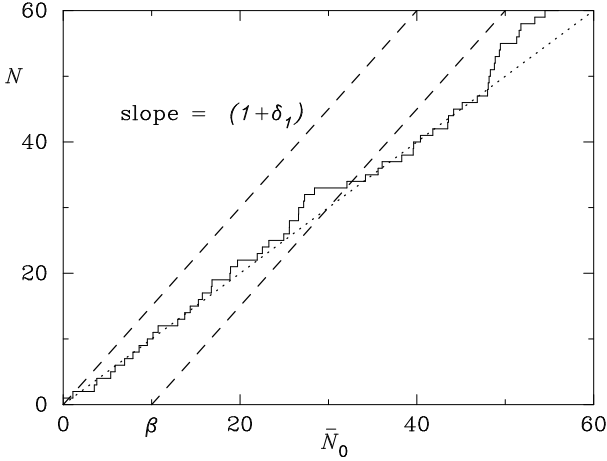}}
\caption{Example of the trajectory (solid line) traced by the number, 
$N$, of Poisson--distributed objects within a sphere which contains 
$\bar N_0$ objects on average.  The dotted line, which has unit slope, 
shows the curve traced out by the average density.  The dashed line 
which starts at the origin shows the barrier considered by 
Epstein (1983) and Sheth (1995b).  The dashed line which starts at 
$\bar N_0 = \beta$ is the shifted barrier studied here.  }
\label{trajec}
\end{figure}

Since $f_{\rm c}(j,b_1)$ is a statement about the last crossing 
of the barrier (the dashed line that intersects the origin 
in Fig.~\ref{trajec}) by the random walk, excursion set trajectories, 
it will sometimes be referred to as the barrier crossing 
distribution.  The subscript `c' here denotes the fact that 
this is the distribution of last crossings of a constant boundary; 
that is, $\delta_{\rm c0}$ is independent of $V_0$.  

Let $f_{\rm c}(j,b_1|N,b_2)$ denote the fraction of 
trajectories for which $V_0(j)$ is the largest value of $V_0$ 
at which $N(V_0)\ge \bar N_0/b_1$, given that, at some 
$V_0'\equiv V_0(N)>V_0(j)$ they have value $N(V'_0) = \bar N_0/b_2$, 
with $b_2\ge b_1$, and that all $V_0>V_0(N)$ are less dense 
than $V_0(N)$.  Then 
\begin{eqnarray}
f_{\rm c}(j,b_1|N,b_2) &=& N\left(1-{b_1\over b_2}\right)\ 
{N\choose j}\ {j^j\over N^N}\ 
\left({b_1\over b_2}\right)^{j-1} \nonumber \\
&&\qquad \ \ \times\ \left(N - {jb_1\over b_2}\right)^{N-j-1}
\label{fjk}
\end{eqnarray}
(Sheth 1995b).  

It is usual to associate these expressions about the statistics 
of trajectories crossing a constant barrier with statements about 
the number density of collapsed (point--sized) halos.  Thus, the 
average number density of $b_1$-halos that contain exactly $j$ 
particles is 
\begin{equation}
\bar n_0(j,b_1) = \bar n_0\,{f_{\rm c}(j,b_1)\over j} = 
\bar n(1-b_1)\,\eta(j,b_1).
\label{njfj}
\end{equation}
This assignment comes from assuming that the fraction of 
trajectories $f_{\rm c}(j,b_1)$ can be equated to the fraction 
of the Lagrangian space associated with regions containing $j$ 
particles with overdensity $b_1$.  The final equality comes 
from equation~(\ref{fjb}) and using the fact that 
$\bar n_0\equiv\bar n$.  

Similarly, the average number of $(j,b_1)$-halos that are 
within an $(N,b_2)$-halo is 
\begin{equation}
{\cal N}(j,b_1|N,b_2) = 
\left({N\over j}\right)\,f_{\rm c}(j,b_1|N,b_2),
\label{nj1k2}
\end{equation}
where $N\ge j$ and $b_2\ge b_1$.  

If the trajectories are not centred on particles, then the 
barrier crossing distribution is 
\begin{equation}
F_{\rm c}(j,b_1) = b_1\ f_{\rm c}(j,b_1),
\end{equation}
and 
\begin{equation}
F_{\rm c}(j,b_1|N,b_2) = (b_1/b_2)\ f_{\rm c}(j,b_1|N,b_2).
\end{equation}
However, the number density of associated regions is the same as 
before (Sheth \& Lemson 1998).  

Consider an $(N,b_2)$-halo that is known to contain $m$ $b_1$ 
subhalos, of which $n_1$ are singles, $n_2$ are doubles and so on.  
Thus, $\sum_{j=1}^N n_j = m$, and mass conservation requires that 
$\sum_{j=1}^N j\,n_j = N$.  Of course, $b_1\le b_2$.  
Let $\pi[{\bmath n}|N]$ denote one particular partition of $N$, 
where ${\bmath n}$ denotes the vector $(n_1,\cdots\,,n_N)$, and 
let $p({\bmath n};b_1|N;b_2)$ denote the probability that the 
partition $\pi[{\bmath n}|N]$ occured.  Sheth (1996b) shows that 
\begin{equation}
p({\bmath n};b_1|N;b_2) =
{(Nb_{21})^{m-1}{\rm e}^{-Nb_{21}}\over \eta(N,b_2)}\,
\prod_{j=1}^N {\eta(j,b_1)^{n_j}\over n_j!},
\label{prtfnc}
\end{equation}
where $b_{21}=(b_2-b_1)$, and $Nb_2 = \bar nV_0(N)$, 
is consistent with the excursion set model described above 
(also see Sheth \& Pitman 1997, Sheth \& Lemson 1998).  

For example, the average number of $(j,b_1)$-halos that are within 
$(N,b_2)$-halos is 
\begin{equation}
\bigl\langle n_j;b_1|N;b_2\bigr\rangle = \!\!
\sum_{\pi[{\bmath n}|N]}\!\! n_j\,p({\bmath n};b_1|N;b_2) = 
{\cal N}(j,b_1|N,b_2) ,
\label{meanj}
\end{equation}  
and the sum is over the set of all distinct ordered 
partitions of $N$.  The final expression is the same as 
equation~(\ref{nj1k2}) as required;  the algebra leading to it is 
given in Appendix B of Sheth (1996b).  

The correlation between $(i,b_1)$- and $(j,b_1)$-halos that are 
within the same $(N,b_2)$-halo is computed by a similar average 
over all partitions of $N$.  Thus, 
\begin{equation}
\bigl\langle n_i\,n_j;b_1|N;b_2\bigr\rangle = 
{\cal N}(j,b_1|N,b_2)\ {\cal N}(i,b_1|N-j,b'),
\label{ninj}
\end{equation}
where 
\begin{displaymath}
(N-j)b' = Nb_2 - jb_1 
\end{displaymath}
(Sheth 1996b).  This expression reflects the fact that, in the 
Lagrangian Poisson distribution, non-overlapping volumes 
are mutually independent (Sheth \& Lemson 1998). 
These expressions will be useful later.  

This subsection shows how the statistics of the initial 
Lagrangian distribution can be used to derive some 
useful information about the statistics of collapsed halos, 
and of the distribution of halos within halos.  While the language 
of halos is useful, it is important to remember that an 
$(N,b_2)$-halo can also be thought of as a Lagrangian region of 
size $V_0(N) = Nb_2/\bar n$ (Mo \& White 1996).  Thus, expressions 
like~(\ref{ninj}) are related to the average number of $(i,b_1)$- 
and $(j,b_1)$-halos that are both within the same Lagrangian region 
of size $V_0(N)$.  
It is in this sense that many of the expressions above will be 
interpretted later in this paper.  

\subsection{The excursion set with shifted barrier}\label{shift}
The previous subsection considered the distribution of last 
crossings, by random walk trajectories associated with the Poisson 
distribution, of a barrier which had shape $\bar nV_0(j) = jb$.  
Instead, suppose that 
\begin{equation}
\bar nV_0(j) \equiv \beta + jb, \qquad{\rm with}\ 0\le b<1,
\label{newvj}
\end{equation}
and we seek an expression for the fraction of trajectories 
$f(j,b,\beta)$ for which $V_0(j)$ is the largest value of $V_0$ 
at which $N(V_0)=(\bar N_0-\beta)/b$.  This is equivalent to 
considering the same problem as before, but with the barrier 
shifted to the right by a constant $\beta$ (see Fig.~1).  

To compute $f(j,b,\beta)$, start with an arbitrarily small 
sphere centred on a particle.  Since the distribution is Poisson, 
counts in different volumes are independent, so 
\begin{equation}
f(j,b,\beta) = p\Bigl(j-1|V_0(j)\Bigr)\,f^{\rm E}(j,b,\beta) ,
\end{equation}
where $p(j-1|V_0)$ is the Poisson distribution of 
equation~(\ref{pois}), with $V_0(j)$ given by equation~(\ref{newvj}), 
and $f^{\rm E}(j,b,\beta)$ denotes the probability that no sphere 
larger than and concentric to $V_0(j)$ is denser than the threshold 
value.  Now, $f^{\rm E}(j,b,\beta)$ is the same as one minus 
the probability that $V_0(N)$ is the largest volume denser than 
the threshold value, summed over all $N\ge j$.  As a consequence 
of the Poisson assumption, $f^{\rm E}(N,b,\beta)$ is independent 
of $N$, so it can be written as $f^{\rm E}(b,\beta)$ 
(e.g. Epstein 1983).  This means that 
\begin{equation}
1-f^{\rm E}(b,\beta) = f^{\rm E}(b,\beta)
\sum_{N=j}^\infty p\Bigl(N-j|V_0(N)-V_0(j)\Bigr).
\label{fe}
\end{equation}
Since $\bar n[V_0(N)-V_0(j)]=(N-j)b$, setting $m=(N-j)$ means that 
the sum above is simply 
\begin{equation}
\sum_{m=0}^\infty {(mb)^m\,{\rm e}^{-mb}\over m!} = {b\over 1-b}.  
\label{mborel}
\end{equation}
The final expression follows from recognizing that the term 
in the sum is $(mb)$ times the Borel$(b)$ distribution.  
This implies that 
\begin{equation}
f^{\rm E}(j,b,\beta) = f^{\rm E}(b,\beta) = (1-b),  
\label{fext}
\end{equation}
so that 
\begin{equation}
f(N,b,\beta) = (1-b)\ {(\beta + Nb)^{N-1}\over (N-1)!}\ 
{\rm e}^{-\beta - Nb}.
\label{fnb}
\end{equation}
This is the barrier crossing distribution associated with the 
shifted barrier.  Following equation~(\ref{njfj}), this barrier 
crossing distribution can be associated with a number density 
of Lagrangian regions which contain $N$ particles:
\begin{equation}
\bar n(N,b,\beta) = \bar n\ {f(N,b,\beta)\over N}.  
\label{nlgrng}
\end{equation} 

Let $F(N,b,\beta)$ denote the barrier crossing distribution if the 
trajectory is centred on a random position, not necessarily 
on a particle.  Then 
\begin{equation}
F(N,b,\beta) = (1-b)\,p\Bigl(N|V_0(N)\Bigr) 
= {\bar nV_0(N)\over N}\,f(N,b,\beta).  
\end{equation}
The number density of associated regions, $\bar n(N,b,\beta)$, 
is related to this fraction analogously to how it is related 
to $f(N,b,\beta)$.  Namely, the barrier crossing distribution 
should be weighted by the number of trajectories associated with it, 
so $\bar n(N,b,\beta)$ is $F(N,b,\beta)$ times the ratio of the 
average density $\bar n$ to $\bar nV_0(N)$, so it is given by
equation~(\ref{nlgrng}).

\subsection{Statistics in the Eulerian space}\label{expdf}
Imagine partitioning the Eulerian space $V_{\rm tot}$ containing 
$N_{\rm tot}$ particles into a large number of cells, each of 
size $V$.  Then the total number of such cells is $V_{\rm tot}/V$.  
We will be interested in the limit in which 
$N_{\rm tot}/V_{\rm tot}\to \bar n$ as both 
$N_{\rm tot}\to\infty$ and $V_{\rm tot}\to\infty$.  
Let $p(N|V)$ denote the fraction of these cells that contain 
exactly $N$ particles.  If $n(N|V)$ denotes the number of 
cells containing exactly $N$ particles, then 
\begin{equation}
p(N|V) \equiv {n(N|V)\over N_{\rm tot}}
= {V\,n(N|V)\over V_{\rm tot}} 
\label{pmr}
\end{equation}
is said to be the Eulerian counts-in-cells distribution.  

Suppose that $\bar n$ and $V_{\rm tot}$ in the Lagrangian and 
Eulerian spaces are the same.  Then $\bar n_0 = \bar n$.  
Further, suppose that the number of regions which contain a 
specified number of particles is the same in both the Lagrangian 
and the Eulerian spaces.  Finally, suppose that the parameter 
$\beta$, which controlled the shape of the barrier of the previous 
subsection, can be related to the Eulerian cell size $V$.  
Then equation~(\ref{nlgrng}) requires that 
\begin{equation}
{V\,n(N|V)\over V_{\rm tot}} \equiv V\,\bar n(N|V) = 
{\bar nV\,f(N|V)\over N},
\label{conserve}
\end{equation}
so that 
\begin{equation}
p(N|V) = {f(N|V)\over N/\bar N},\qquad{\rm where}\ 
\bar N\equiv \bar nV.
\label{fspd}
\end{equation}
Equation~(\ref{fspd}) provides a relation between the barrier 
crossing distribution $f(N|V)$, which itself depends on the 
shape of the boundary and the initial Lagrangian field, and the 
Eulerian counts-in-cells distribution.  By hypothesis, the 
shape of the boundary depends on the Eulerian scale $V$, so 
a given relation between $\beta$ and $V$ implies a specific form 
for the evolution of the comoving sizes of regions.  This is 
discussed in more detail in Section~\ref{scm}.  
Physically, equation~(\ref{fspd}) is consistent with the assumption 
that the difference between the particle distribution in the 
initial (Lagrangian) and final (Eulerian) spaces arises solely 
as a consequence of the fact that, although the comoving sizes 
of regions may change, the number of expanding and contracting 
regions in the two spaces is conserved.  

Let $\Delta\equiv(1+\delta)\equiv N/\bar N$.  Then, 
$p(\Delta|V)$ is the probability distribution function 
of the density in Eulerian space, and 
\begin{equation}
\int_0^\infty \!\!p(\Delta|V)\ {\rm d}\Delta = 
\int_0^\infty \!\!\Delta\ p(\Delta|V)\ {\rm d}\Delta = 1.
\label{norm}
\end{equation}

Following equation~(\ref{fspd}), equation~(\ref{fnb}) has an 
associated Eulerian counts in cells distribution:
\begin{eqnarray}
p(N|V,b,\beta) &\equiv& {f(N,b,\beta)\over N/\bar N} \nonumber\\
&=& {\bar N(1-b)\over N!}\ (\beta + Nb)^{N-1}\,
{\rm e}^{-\beta - Nb} ,
\label{gpd}
\end{eqnarray} 
where $\bar N\equiv\bar nV$.  This is the Generalized Poisson 
distribution (equation~\ref{gpdf}).  Normalization to unity 
requires that 
\begin{equation}
\beta = \bar N(1-b),\qquad{\rm where}\ \bar N\equiv\bar nV,
\label{betab}
\end{equation}
so the variance is $\bar N/(1-b)^2$, and 
this distribution is the same as that in equation~(\ref{gpdf}).  

Equation~(\ref{betab}) shows how the parameter $\beta$ is related 
to the Eulerian cell size $V$.  In particular, notice that as 
$V\to 0$, then the barrier is shifted by $\beta\to 0$, so, in this 
limit, the barrier shape is the same as that in Section~\ref{const}.  
This shows explicitly that, as $V\to 0$, the Eulerian counts in 
cells distribution is the same as the halo mass function.  

\subsection{The halo distribution}\label{halos}
Recall that the Generalized Poisson distribution with parameter 
$b$ can be interpretted as arising from a Poisson distribution 
of Borel$(b)$ halos.  This subsection shows that the derivation 
of the Generalized Poisson distribution presented earlier is 
consistent with this interpretation.  

Fig.~\ref{trajec} shows that the fraction of trajectories which 
last cross the constant barrier, parameterized by $b_1$, at $j$ 
is equal to the fraction of those trajectories which last crossed 
the shifted barrier (associated with the Eulerian scale $V$ and 
parameter $b\ge b_1$) at $N\ge j$, that had their last crossing 
of the constant barrier at $j$, summed over all $N\ge j$.  
A little algebra shows that 
\begin{equation}
f_{\rm c}(j,b_1) = 
\sum_{N=j}^\infty f_{\rm c}(j,b_1|N,b_2)\,f(N,b,\beta) ,
\ \ {\rm with}\ b\ge b_1,
\end{equation}
where $f_{\rm c}(j,b_1)$ and $f_{\rm c}(j|N)$ are given by 
equations~(\ref{fjb}) and~(\ref{fjk}), and 
$b_2 = \bar N_0/N = (\beta + Nb)/N$, as required by 
equation~(\ref{newvj}).  This relation implies that 
\begin{eqnarray}
\bar n(j,b_1)V &=& 
\sum_{N\ge j}^\infty {\cal N}(j,b_1|N,b_2)\ p(N|V,b,\beta) \nonumber\\
&=& \bar n_0(j,b_1)V ,
\label{same}
\end{eqnarray}
where the final expression is $V$ times equation~(\ref{njfj}), 
and follows from setting $b_2 = (\beta + Nb)/N$.  

There are two reasons for writing this calculation out in detail.  
The first is simply to show explicitly that the mean number density 
of $(j,b_1)$-halos in the Eulerian space is the same as in 
the Lagrangian space, as required.  The second is that it shows 
how statistics in the Lagrangian space can be used to compute 
statistics in the Eulerian space.  Recall that the number of regions  
containing $N$ particles is the same in both spaces, though the 
sizes $V_0$ and $V$ of the regions may be different.  In particular, 
the Lagrangian scale associated with an Eulerian region of size $V$ 
depends on the number of particles $N$ within it:  
$V_0(N) = (\beta + Nb)/\bar n$ (equation~\ref{newvj}).  
So, to compute averages over Eulerian cells $V$, one simply needs 
to sum over the relevant Lagrangian regions $V_0(N)$ that now have 
Eulerian scale $V$, and weight by the Eulerian probability $p(N|V)$ 
that the Eulerian region $V$ contains $N$ particles.  

Thus, the cross-correlation between halos and mass, averaged over 
Eulerian cells $V$, can be computed as follows.  Define 
\begin{equation}
\delta_{\rm h}(j,b_1|N,b_2) = 
{{\cal N}(j,b_1|N,b_2)\over \bar n_0(j,b_2)V} - 1.
\label{dhmow}
\end{equation}
This is the number of $(j,b_1)$-halos that are within 
Lagrangian regions of size $V_0(N)=Nb_2/\bar n$ and which contain 
exactly $N$ particles, divided by the average number of 
$(j,b_1)$-halos that are within Eulerian volumes of size $V$, 
minus one.  By hypothesis, the number of such Lagrangian regions 
is constant, only their size has changed.  However, if we now 
require that $b_2 = (\beta+Nb)/N$, then this means that the 
Eulerian size of such a Lagrangian region is $V$.  So, if we 
require that $b_2 = (\beta+Nb)/N$, then equation~(\ref{dhmow}) 
is the number of $(j,b_1)$-halos in Eulerian cells $V$ that 
contain exactly $N$ particles, relative to the average number of 
$(j,b_1)$-halos in Eulerian cells $V$, minus one.  

The cross correlation function between $(j,b_1)$-halos 
and mass, averaged over all Eulerian cells $V$, is 
$\delta_{\rm h}(j,b_1|N,b_2)$, with $Nb_2 = \beta+Nb$, times 
$\delta \equiv (N-\bar N)/\bar N$ times the probability that an 
Eulerian region $V$ contains exactly $N$ particles, summed over 
all $N$.  Thus, 
\begin{eqnarray}
\bar\xi_{\rm hm}(j,b_1,\beta) &\equiv& 
\Bigl\langle \delta_{\rm h}(j,b_1|N,b_2)\ \delta\Bigr\rangle
\nonumber \\
&=& \sum_{N=j}^\infty \left({N\over\bar N}-1\right)\,
{{\cal N}(j,b_1|N,b_2)\over\bar n(j,b_1)V}\ p(N|V,b,\beta) \nonumber \\
&=& \sum_{N=j}^\infty 
\left({N\over\bar N}-1\right)\,
{f_{\rm c}(j,b_1|N,b_2)\over f_{\rm c}(j,b_1)}\ 
f(N,b,\beta) \nonumber \\
&=& {j\over \bar N}\left({1-b_1\over 1-b}\right) + 
{(b-b_1)\over \bar N(1-b)^2(1-b_1)},
\end{eqnarray}
where ${\cal N}(j|N)$ is given by equation~(\ref{nj1k2}), 
$\bar n(j,b_1)$ by equation~(\ref{same}), and $p(N|V,b,\beta)$ by 
equation~(\ref{gpd}).  
The second line follows from equation~(\ref{dhmow}), and the fact 
that $\langle\Delta\rangle = \langle 1+\delta\rangle = 1$ 
(equation~\ref{norm}), so $\langle\delta\rangle=0$.  
The third line follows from equations~(\ref{njfj}), (\ref{nj1k2}) 
and~(\ref{fspd}), and the last line from doing the sum, after 
using the fact that $Nb_2 = (\beta+Nb)$.  
Notice that when $b=b_1$ then $\bar\xi_{\rm hm} = j/\bar N$.  

The correlation between $b_1$-halos of mass $i$ and $j$, 
averaged over Eulerian cells $V$, arises as a result of two 
averages.  The first is over all possible ways the $N$ particles 
in an Eulerian cell $V$ could have been partitioned into $b_1$-halos, 
and the second is over all possible values of $N$.  The assumption that 
an Eulerian cell $V$ is simply a Lagrangian region that has changed 
size allows us to assume that the first average (over all partitions 
of $N$) is the same in the two spaces.  So, the result of this 
average is $\langle n_i\,n_j;b_1|N;b_2\rangle$ of 
equation~(\ref{ninj}), provided we set $b_2 = (\beta+Nb)/N$.  
All that remains is to average this quantity over $N$, and 
then divide out the factors expected on average.  Thus, 
\begin{equation}
\bar\xi_{\rm hh}(i,j,b_1|V) =\!\!\! \sum_{N=i+j}^\infty \!\!
{\bigl\langle n_i\,n_j;b_1|N;b_2 \bigr\rangle \over 
\bar n(i,b_1)V\,\bar n(j,b_1)V}\,p(N|V,b,\beta) - 1,
\end{equation}
where 
\begin{displaymath}
Nb_2 = \beta + Nb, \qquad{\rm and}\ \ (N-j)b' = Nb_2 - jb_1 .
\end{displaymath}
This sum can be solved analytically:  
\begin{equation}
\bar\xi_{\rm hh}(i,j,b_1|V) = {(i+j)(b-b_1)\over\bar N(1-b)} + 
{(b-b_1)^2\over\bar N(1-b)^2(1-b_1)^2}.
\end{equation}

When $b=b_1$, $\bar\xi_{\rm hh}=0$, for all $i,j$, and $V$, which 
implies that the halos have a Poisson distribution.  This is 
consistent with the fact that equation~(\ref{gpd}) is a Compound 
Poisson distribution which arises if Borel$(b)$ clusters have a 
Poisson spatial distribution (Saslaw 1989; Sheth \& Saslaw 1994).  
When $b_1=0$, then all halos are single particles, so $i=j=1$, 
and this expression gives the second factorial moment of the 
single particle distribution.  Simple algebra shows that, in 
this limit, it is equal to the second factorial moment of 
equation~(\ref{gpd}).  Further, notice that when $b_1\le b$, 
then correlations between halos only depend on the sum of the 
halo masses, not on the masses of the individual halos themselves.  
This suggests that, in this model, halo-halo correlations arise 
because of volume exclusion effects only.  That is, halo-halo 
correlations arise only because, initially, a $(j,b_1)$-halo 
occupies a region $V_0(j)=jb_1/\bar n$, so no other halos can 
occupy this region.  As time passes, such an object collapses to 
a region of zero Eulerian size, so the volume excluded by it 
becomes negligibly small.  

\begin{figure}
\centering
\mbox{\psfig{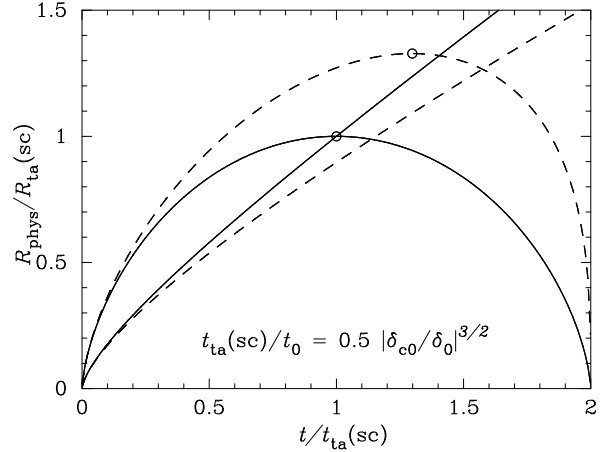}}
\caption{The physical radius of a perturbation in units of the 
spherical model turnaround radius, as a function of time in 
units of the spherical model turnaround time.  The solid curves 
show the spherical model, and dashed curves show the model 
developed here.  The two curves for each line type are for 
denser perturbations (which recollapse) and under-dense 
perturbations (which do not).}
\label{rphys}
\end{figure}

\subsection{Relation to the spherical collapse model}\label{scm}
In the spherical Poisson model, $b$ is related to the 
critical overdensity required for collapse:  
$b = 1/(1+\delta_{\rm c0})$ (equation~\ref{vj}).  
This relation for $b$, with equation~(\ref{betab}) 
for $\beta$ and equation~(\ref{newvj}), imply that, 
as $N$ changes, the height relative to the average density 
of the shifted barrier considered here is 
\begin{equation}
\delta_0(N|V) \equiv {N\over \bar N_0} - 1 = 
{(\bar N_0-\beta)\over \bar N_0b} - 1 = 
\delta_{\rm c0} - {\bar N\over \bar N_0}{(1-b)\over b} .
\end{equation}
When $b\to 1$, and $N\gg 1$, then $\delta_{\rm c0}\ll 1$ and 
$\bar N/\bar N_0\to \bar N/N$, so 
\begin{equation}
\delta_0 \to \delta_{\rm c0} - {\delta_{\rm c0}\over 1+\delta}.
\label{linbar}
\end{equation}
In this limit, the relation between $\delta_0$ and $\delta$ 
is independent of $V$.  This relation should be compared with 
that for the spherical collapse model given in Appendix~\ref{scoll}.

In the limits $\delta_{\rm c0}\ll 1$ and $N\gg 1$, 
equation~(\ref{linbar}) suggests the following model for the 
collapse of objects.  
Let $R(z)$ denote the comoving size of an object at the epoch $z$.  
Then $R(z)=R_0$ initially.  If the object is in an underdense 
region, then its comoving size will increase, else it will 
decrease.  Trajectories with extrapolated linear overdensity 
$\delta_0$ greater than $\delta_{\rm c0}$ are associated with 
collapsed objects.  Collapsed objects have $R(z)=0$.  Until 
collapse 
\begin{equation}
{V(z)\over V_0} = {R^3(z)\over R_0^3} = 
1 - {\delta_0/(1+z)\over \delta_{\rm c0}}.
\end{equation}
The radius of an object in proper, physical coordinates 
is $R_{\rm p}(z) = R(z)/(1+z)$.  Objects which collapse have 
a turnaround radius--the maximum value that $R_{\rm p}(z)$ 
attains.  This occurs at 
\begin{equation}
(1+z_{\rm ta}) = {4\over 3}{\delta_0\over\delta_{\rm c0}} ,
\end{equation}
at which time 
\begin{equation}
{R(z_{\rm ta})\over R_0} = {1\over 4^{1/3}}.
\end{equation}
Figure~\ref{rphys} shows that, for overdense perturbations, 
turnaround in this model (dashed lines) occurs later, and at a 
larger radius, than in the spherical model (solid lines).  
In contrast, underdense regions expand less rapidly in this model 
than in the spherical model.  

\section{The associated queue}\label{queue}
The excursion set problem studied above can be understood in 
terms of a simple queue system.  
A single server queue with deterministic service time $0\le b\le 1$ 
and Poisson arrivals with unit mean, which starts with one customer 
at the initial time, can be expressed in terms of the random walk 
problem studied by Epstein (1983) and Sheth (1995b).  The time 
parameter in the queue system is like the cell size parameter in 
the excursion set model.  

Consider the probability $B(j,b)$ that, 
after exactly $j$ customers have been served, the queue is empty 
for the first time.  Then $f(j,b)$ of equation~(\ref{fjb}) is $j$ 
times this probability times $(1-b)$.  The $(1-b)$ factor simply 
comes from the additional constraint in the excursion problem on 
the number of particles within volumes larger than the critical 
volume $V_0(j)$.  This constraint is not present in the queue 
model, since we have made no assumption about what happens in the 
queue after the end of the first busy period.  

Figure~\ref{trajec} shows clearly that the same queue system, 
started with $m$ customers at the initial time, is related to the 
excursion set problem studied in the previous section.  Let 
$B(N,b|m)$ denote the probability that exactly $N$ customers were 
served in the first busy period, given that there were exactly $m$ 
customers in the queue initially.  Tanner (1953, 1961) shows that 
\begin{equation}
B(N,b|m) = {m\over N}\,{(Nb)^{N-m}\,{\rm e}^{-Nb}\over (N-m)!},
\end{equation}
and he also discusses the origin of the $(m/N)$ term.  
Notice that $B(N,b|1)$ is the Borel$(b)$ distribution.  

The excursion set probability $f(N,b,\beta)$ of equation~(\ref{fnb}) 
is the same as the probability that there were exactly $m$ customers 
at the start, times $(N/m)\,B(N,b|m)$, times $(1-b)$, summed over 
all possible values of $m$.  
In the excursion set problem, the probability that there are 
exactly $m$ customers at the start is simply 
\begin{equation}
p(m,\beta) = {\beta^{m-1}\,{\rm e}^{-\beta}\over (m-1)!}
\end{equation}
(Figure~\ref{trajec}), so that 
\begin{equation}
f(N,b,\beta) = \sum_{m=0}^N (1-b)\,(N/m)\,B(N,b|m)\,p(m,\beta).
\end{equation}
Appendix~D of Sheth (1996b) shows that the right hand side of this 
expression is the same as that on the right hand side 
equation~(\ref{fnb}).  The corresponding expression for 
trajectories not necessarily centred on a particle, 
$F(N,b,\beta)$, can also be derived in this context.  Simply set 
$p(m,\beta)$ to $(\beta/m)$ times the expression above, and do 
the sum.  Sheth (1996b) discusses a branching process 
derivation of this formula.  Thus, this section shows how that 
branching process, this queue model, and the excursion set model 
of the previous section are all interrelated.  

\section{A scaling solution}\label{scale}
This subsection extends the results of Sheth (1995b), Section~3, 
for a constant barrier to the shifted barrier considered in 
this paper.  In particular, it shows that the shifted barrier problem 
has a scaling solution that is analogous to the one associated with 
the constant barrier.  

Suppose that the underlying distribution is not Poisson, but is 
Compound Poisson.  This means that equation~(\ref{pois}) must be 
replaced with $P_{\rm CP}[n|V_0(n)]$.  

Then 
\begin{equation}
f_{\rm CP}(n,\delta) = f_{\rm CP}^{\rm I}(n,\delta) \ 
f_{\rm CP}^{\rm E}(n,\delta) ,
\end{equation}
where the first term is the probability that there are exactly $n$ 
particles within $V_0(n)$, and the second is the probability that 
all volumes larger than $V_0(n)$ are less dense than it.  The 
second term is obtained by noting that the argument leading to 
equation~(\ref{fext}) holds here also.  Therefore, when 
$P_{\rm CP}(n|V_0,b_0)$ is the Generalized Poisson distribution 
with parameter $b_0$, and $V_0(n)$ is given by 
equation~(\ref{newvj}), i.e., 
\begin{displaymath}
\bar nV_0(n) \equiv \beta + nb = \bar N(1-b) + nb \equiv\bar N_n,
\end{displaymath}
then $f_{\rm GPD}^{\rm E}$ is given by an expression 
like~(\ref{fe}) but with $p$ replaced by $P_{\rm CP}$.  
Since $\bar n[V_0(n)-V_0(j)] = (n-j)b$, setting $m\equiv n-j$ 
and 
\begin{equation}
B = b_0 + b\,(1-b_0) 
\end{equation}
means that 
\begin{equation}
{1\over f_{\rm GPD}^{\rm E}} = 
1 + \sum_{m=0}^\infty mb(1-b_0)\,{(mB)^{m-1}{\rm e}^{-mB}\over m!}.
\end{equation}
This sum is similar to that in equation~(\ref{mborel}).  Thus, 
\begin{equation}
f_{\rm GPD}^{\rm E} = \left(1 + {b(1-b_0)\over 1-B}\right)^{-1} = 1 - b.  
\end{equation}

The other term is slightly more complicated.  Recall that 
the Generalized Poisson distribution with parameter $b_0$ can 
be understood as describing a Poisson distribution of point--sized 
clusters, where $\eta(m,b_0)$, the probability that a randomly chosen 
cluster contains exactly $m$ particles is the Borel$(b_0)$ 
distribution.  Since the mean of the Borel$(b_0)$ distribution 
is $(1-b_0)^{-1}$, the ratio of the density of cluster centres 
to that of particles is $\bar n_{\rm clus}/\bar n = (1-b_0)$, and 
\begin{eqnarray}
f_{\rm GPD}^{\rm I}(n) &=& \sum_{m=1}^n 
{\bar n_{\rm clus}\over \bar n}\,m\,\eta(m,b_0)\,\nonumber \\
&&\qquad\qquad \times\ \ P_{\rm GPD}\Bigl(n-m\Bigl\vert V_0(n),b_0\Bigr) 
\end{eqnarray}
(Sheth 1995b equation~26).  
Abel's generalization of the Binomial formula (e.g. Sheth 1995b 
equation~30) reduces this to 
\begin{equation}
f_{\rm GPD}(n) = {(1-B)\over (n-1)!}\ 
\Bigl[\theta + nB\Bigr]^{n-1}\ 
{\rm e}^{-\theta - nB} ,
\end{equation}
where 
\begin{displaymath}
1-B \equiv (1-b)(1-b_0) \qquad{\rm and}\qquad 
\theta \equiv \bar N(1-B).
\end{displaymath}
This has the same form as equation~(\ref{fnb}).  Notice that when 
$b_0=0$, then $B=b$, and this expression is exactly the same as 
equation~(\ref{fnb}).  This is sensible, since a Generalized 
Poisson distribution with parameter $b_0=0$ is just a 
Poisson distribution.  

Since $b$ is a pseudo-time variable, the new definition of 
$B$ simply means that the time parameter in this model is 
slightly different from that in the case of Poisson 
initial conditions.  
In other words, if the initial Lagrangian distribution is 
Generalized Poisson, rather than Poisson, then, except for 
the appropriate rescalings of the time-dependent parameters, 
none of the results of Section~2 are changed.  

\section{The two barrier problem}\label{twobar}
Suppose $b_1\le b_2$.  Let $f(j,b_1|k,b_2)$ denote the fraction 
of trajectories, centred on a particle, which have $j$ particles 
when they last crossed the barrier with index $b_1$, when it is 
known that they have exactly $k$ particles when they last cross 
the barrier with index $b_2$.  When $k\ge \bar N$, then the 
results of Sheth (1995b) imply that 
\begin{eqnarray}
f(j,b_1|k,b_2) &=& {k-1\choose j-1} 
\,{{_2\!V_k} -  {_1\!V_k}\over _2\!V_k^{k-1}}\ \nonumber \\
&&\qquad \times\ \ (_1\!V_j)^{j-1}\,(_2\!V_k - {_1\!V_j})^{k-j-1} ,
\label{fj1k0}
\end{eqnarray}
where $j\le k$, 
\begin{displaymath}
_1\!V_j = \bar N(1-b_1) + jb_1,\qquad _1\!V_k = \bar N(1-b_1) + kb_1, 
\end{displaymath}
\begin{displaymath}
{\rm and}\  _2\!V_k = \bar N(1-b_2) + jb_2 .
\end{displaymath}
(Notice that equation~\ref{fjk} is this expression with 
$_1\!V_j=jb_1$, $_1\!V_k=kb_1$, and $_2\!V_k=kb_2$.)  
This reflects that fact that a comoving volume which is denser 
than the average density at time $b_2$ will have been less dense 
at an earlier time.  When $k<\bar N$, then 
\begin{eqnarray}
f(k,b_2|j,b_1) &=& {j-1\choose k-1} 
\,{_1\!V_j - {_2\!V_j}\over _1\!V_j^{j-1}}\nonumber \\
&&\qquad\times\ \ (_2\!V_k)^{k-1}\,(_1\!V_j - {_2\!V_k})^{j-k-1} ,
\end{eqnarray}
but now $k\le j$, since a comoving volume that is less dense 
than the average at some late time $b_2$ must also have been 
underdense at the earlier time $b_1\le b_2$, and its density 
will have decreased since the earlier time.  This expression 
is the probability that a cell contains exactly $k$ particles 
at time $b_2$ given that at some earlier time $b_1\le b_2$ it 
contained exactly $j$ particles.  These expressions are the 
analogues of equation~(\ref{fjk}).  

In the limits $k\gg \bar N$ and $k\gg j$, the use of 
Stirling's approximation shows that 
\begin{equation}
f(j|k) \to f(j,B,\beta') ,
\end{equation}
where $f(j,B,\beta')$ has the same form as equation~(\ref{fnb}) 
with 
\begin{equation}
B = {kb_1\over \bar n{_2\!V_k}}\qquad{\rm and}\ \ 
\beta' = {b_1\ \bar N(1-b_1)\over \bar n{_2\!V_k}}.
\end{equation}
This is similar to the rescaling associated with the constant 
barrier:  When $N\gg j$ and $b_2\ge b_1$, then 
\begin{equation}
f_{\rm c}(j,b_1|N,b_2) \to f_{\rm c}(j,b_1/b_2).  
\end{equation}
This rescaling, and the scaling solution of the previous section, 
suggest that there may be a merger--fragmentation model of the 
type described by Sheth \& Pitman (1997) associated with the 
Generalized Poisson distribution.  

For trajectories that are not necessarily centred on particles, 
the expression corresponding to equation~(\ref{fj1k0}) is 
\begin{equation}
F(j,b_1|k,b_2) = {k\choose j} 
\,{{_2\!V_k} -  {_1\!V_k}\over ({_2\!V_k})^k}\ 
(_1\!V_j)^j\,({_2\!V_k} - {_1\!V_j})^{k-j-1} ,
\end{equation}
when $k\ge \bar N$, and $F(k|j)$ is related to $f(k|j)$ similarly.  
These expressions follow from arguments given in Sheth \& 
Lemson (1998).  Identities associated with Abel's generalization 
of the Binomial theorem show that all these expressions are 
normalized to unity.  

This last expression is related to the following problem.  
Choose a random Eulerian cell of comoving size $V$ in an $N$-body 
simulation, and study the evolution of the mass within it.  
Let $p(j,b_1|k,b_2)$ denote the probability that at time $b_1$ 
there are exactly $j$ particles within it, given that at some 
time $b_2\ge b_1$ it is known to contain exactly $k\ge j$ 
particles.  
Then $p(j,b_1|k,b_2) = F(j,b_1|k,b_2)$.  
These expressions show explicitly that, for some Eulerian cells, 
it may happen that the number of particles within the cell 
decreases initially and increases later.  In other words, in the 
model, matter can flow in and out of Eulerian cells.  

\section{Discussion}
This paper presents a new derivation of the Generalized Poisson 
distribution.  The derivation allows one to construct a useful 
model of hierarchical clustering from Poisson initial conditions.  
The resulting model is useful because the Poisson assumption 
allows one to solve many problems that, at present, have no 
solution if more realistic initial conditions are used.  

The model is a simple generalization of the excursion set model 
developed by Bond et al. (1991).  Their approach allows one to  
estimate the mass function of collapsed halos; the generalization 
presented here allows one to describe the spatial distribution 
of these halos as well.  
The model can also be thought of as a simple variant of the 
spherical collapse model.  In the model, initially 
denser regions contract more rapidly than less dense regions, 
sufficiently underdense regions expand, the influence of 
external tides on the evolution of such regions is ignored, 
and the number of expanding and contracting regions is assumed 
to be conserved.  Strictly speaking, none of these assumptions 
can be justified physically.  However, these simplifications 
mean that the model can be worked out relatively easily. 
Moreover, the Generalized Poisson distribution, derived after 
making these assumptions, is a reasonably accurate fit to the 
Eulerian counts-in-cells distribution measured in numerical 
simulations of clustering from Poisson initial conditions.  
This suggests that, at least for estimating the evolution of 
the counts-in-cells statistic from such initial conditions, 
these simplifications are justified.  

In the model, a collapsed halo occupies a vanishingly small 
volume.  In the simulations, collapsed halos have non-zero 
sizes---any given halo virializes at some fraction, typically 
about one half, of its turnaround radius.  
This means that on scales smaller than that of a typical halo, 
the counts-in-cells distribution computed here will cease to be 
a good approximation to that measured in the simulations.  
As discussed in the introduction, the fact that halos have 
non-trivial density profiles means that the $b$ parameter in 
equation~(\ref{gpdf}) depends on scale.  A reasonable approximation 
to the effects of this scale dependence can be computed from models, 
such as those proposed by Navarro, Frenk \& White (1996), of the 
density profiles of collapsed halos (see Sheth \& Saslaw 1994 for 
details).  As Poisson initial conditions are not realistic anyway, 
this seems an unnecessary refinement to an already idealized model.  

As the basic model has worked out so easily, as it allows 
one to estimate the extent to which halos are biased tracers of 
the mass, and, most importantly, as it provides a reasonably 
accurate description of the evolution of clustering measured in 
numerical simulations, it seems worth extending it to describe 
clustering from more general initial conditions.  
This extension is in progress.  

\section*{Acknowledgments}
I thank Houjon Mo and Simon White for stimulating discussions.

\appendix

\section{The spherical collapse model}\label{scoll}
Consider a region of initial, comoving Lagrangian size $R_0$.  
Let $\delta_0$ denote the extrapolated linear overdensity of 
this object.  In units where the average comoving density is 
unity, there is a deterministic relation between the mass 
$M_0$ within $R_0$: $M_0\propto R_0^3$ provided $\delta_0\ll 1$.  
As the Universe evolves, the size of this region changes.  
Let $R$ denote the size of the region at some later time.  
Then the density within the region is simply 
$(R_0/R)^3\equiv (1+\delta)$.  In the spherical collapse 
model there is a deterministic relation between the initial 
Lagrangian size $R_0$ and density of an object, and its 
Eulerian size $R$ at any subsequent time.  For an 
Einstein--de Sitter universe, 
\begin{eqnarray}
{R_{\rm p}(z)\over R_0} &=& {3\over 10}{1-\cos\theta\over |\delta_0|}
\nonumber \\
{1\over 1+z} &=& {3\times 6^{2/3}\over 20}
{(\theta-\sin\theta)^{2/3}\over |\delta_0|} 
\end{eqnarray}
(e.g. Peebles 1980).  
If $\delta<0$, $(1-\cos\theta)$ should be replaced with 
$(\cosh\theta-1)$ and $(\theta-\sin\theta)$ with 
$(\sinh\theta-\theta)$.  
In this model, collapsing objects reach turnaround at 
\begin{equation}
(1+z_{\rm ta}) = 4^{1/3}\,{\delta_0\over\delta_{\rm c0}},
\end{equation}
at which time 
\begin{equation}
{R(z_{\rm ta})\over R_0} = 
{(1+z_{\rm ta})\,R_{\rm p}(z_{\rm ta})\over R_0} = 
{6\over 10}{4^{1/3}\over \delta_{\rm c0}}.
\end{equation}  

For simplicity, consider the epoch when $z=0$.  
Since $(1+\delta)=(R/R_0)^3$, this means that there is a 
relation between $\delta_0$ and $(1+\delta)$ that is independent 
of $R$.  Mo \& White (1996) give the following approximation 
to this relation:  
\begin{eqnarray}
\delta_0 &=& 1.68647 - 1.35(1+\delta)^{-2/3} - 1.12431(1+\delta)^{-1/2}
\nonumber \\
&&\qquad\qquad +\quad 0.78785(1+\delta)^{-0.58661} .
\label{mow}
\end{eqnarray}
These relations imply that to every pair $(R,z)$ there 
is an associated curve in the $(\delta_0,R_0)$ plane, 
so there is a corresponding curve in the $(\delta_0,V_0)$ plane.   
For a given Eulerian scale $R$, and a specified 
epoch $z$, equation~(\ref{mow}) gives what is effectively the 
boundary $\delta_{\rm sc}(V_0|R)$ associated with the spherical 
collapse model.  This barrier should be compared with that given 
by equation~(\ref{linbar}).  


\begin{thebibliography}{}



\bibitem{bo42} Borel E.,  1942, Compt. rend., 214, 452

\bibitem{co89} Consul P. C.,  1989, Generalized Poisson Distributions:  
Properties and Applications.  M. Dekker, New York 

\bibitem{ep83} Epstein R. I., 1983, MNRAS, 205, 207

\bibitem{iis} Itoh M., Inagaki S., Saslaw W. C., 1993, ApJ, 403, 459



 
\bibitem{mw96} Mo H. J., White S. D. M., 1996, MNRAS, 282, 347

\bibitem{nfw} Navarro J., Frenk C., White S. D. M., 1996, ApJ, 462, 563

\bibitem{pr80} Peebles P. J. E.,  1980, 
The Large Scale Structure of the Universe.  
Princeton Univ. Press, Princeton


\bibitem{wcs89} Saslaw W. C.,  1989, ApJ, 341, 588

\bibitem{sh84} Saslaw W. C., Hamilton A. J. S.,  1984, ApJ, 276, 13

\bibitem{rs95a}  Sheth R. K.,  1995a, MNRAS, 274, 213

\bibitem{rs95b}  Sheth R. K.,  1995b, MNRAS, 276, 796

\bibitem{rs96a}  Sheth R. K.,  1996a, MNRAS, 281, 1124

\bibitem{rs96b} Sheth R. K.,  1996b, MNRAS, 281, 1277

\bibitem{sl97} Sheth R. K., Lemson G., 1998, in preparation

\bibitem{sp97} Sheth R. K., Pitman J.,  1997, MNRAS, 289, 66

\bibitem{rs94} Sheth R. K., Saslaw W. C.,  1994, ApJ, 437, 35

\bibitem{ta53} Tanner J. C.,  1953, Biometrika, 40, 58

\bibitem{ta61} Tanner J. C.,  1961, Biometrika, 48, 222

\bibitem{zhan} Zhan Y., 1989, ApJ, 340, 23

\end{thebibliography}
\end{document}